\documentclass[aps,prl,twocolumn,showpacs,amsmath,amssymb,notitlepage]{revtex4-1}
\usepackage{graphicx,graphics,times,bm,bbm,bbold,color,amsfonts,amsthm,dsfont,hyperref,natbib}

\newcommand{\cA}{\mathbb{A}}
\newcommand{\cB}{\mathbb{B}}
\newcommand{\cG}{\mathbb{G}}

\newcommand{\supp}{\mathrm{supp}}
\newcommand{\diam}{\mathrm{diam}}
\newcommand{\dd}{\mathrm{d}}
\newcommand{\ad}{\mathrm{ad}}
\newcommand{\LR}{\mathrm{LR}}
\newcommand{\ch}{\mathrm{ch}}
\newcommand{\ket}[1]{|#1\rangle}
\newcommand{\bra}[1]{\langle#1|}

\newcommand{\ketbra}[2]{|#1\rangle\langle #2|}
\newcommand{\avg}[1]{\langle#1\rangle}
\newcommand{\abs}[1]{\left|#1\right|}
\newcommand{\norm}[1]{\left\Vert #1\right\Vert }
\newcommand{\ignore}[1]{}
\newtheorem*{theorem}{Theorem}

\begin{document}

\title{Lieb-Robinson Bound and Adiabatic Evolution}

\author{M. M. R. Koochakie}
\affiliation{Department of Physics, Sharif University of Technology, Tehran, Iran}
\author{S. Alipour}
\affiliation{Department of Physics, Sharif University of Technology, Tehran, Iran}
\author{A. T. Rezakhani}
\affiliation{Department of Physics, Sharif University of Technology, Tehran, Iran}

\begin{abstract}
We extend the concept of locality to enclose a situation where a tensor-product structure for the Hilbert space is not \textit {a priori} assumed; rather, this locality is related to a given matrix representation of the Hamiltonian associated to the system. As a result, we formulate a Lieb-Robinson-like bound for Hamiltonians local in a given basis. In particular, we employ this bound to obtain alternatively the adiabatic condition, where adiabaticity is naturally ensued from a locality in energy basis and a relatively small Lieb-Robinson bound.   
\end{abstract}

\pacs{03.65.-w, 03.67.-a, 03.67.Lx, 03.65.Ud}

\maketitle

\textit{Introduction.}---Correlations are responsible for a host of interesting physical phenomena in classical and quantum physics. In particular, in manybody systems where quantum effects prevail, quantum correlations underlie overall physical behavior of the system. A general tool to analyze how quantum correlations spread in systems with \textit{spatially} local Hamiltonians is the Lieb-Robinson (LR) bound \cite{lieb-robinson:CMP,hastings:LesHouches,nachtergaele-sims:arXiv}. This bound dictates a \textit{speed} ($V_\LR$), determined by interaction Hamiltonian/geometry, for how fast observables on a part of a systems (typically a spin lattice) can affect observables on a different (i.e., spatially far) part. In fact, this bound implies an effective \textit{light cone} beyond which information disappears exponentially. Recently, the LR bound has attracted a renewed attention due to its deep implications in quantum manybody theory \cite{hastings-LSM:PRB,hastings:PRL-decay,hastings-wen:PRB,hasting-koma:CMP,nachtergaele-ogata-sims:JSM,nachtergaele-sims:CMP,bravyi-hastings-verstraete:PRL,osborne:PRL,osborne:PRA-gs-1d-approx,osborne:PRA-ad,hastings:JSM-goldstone,hamma:PRL,bravyi-hastings-michalakis:JMP,eisert-cramer-plenio:RMP-area-law,poulin:PRL,barthel:PRL}, quantum information science \cite{osborne:PRA-gapped,keating-etal:PRA,hastings:PRL-ad,burrell-eisert-osborne:PRA,eisert-gross:PRL,wootton-pachos:PRL,dasilva-poulin:tomography,harrison-et-al:arXiv}, and even in mathematics \cite{hastings:CMP-comm-mat,hamma:PRA-generalization-LRB}. Interestingly as well, the LR-like spread of correlations has also been observed experimentally in ultracold bosonic atoms in an optical lattice \cite{LR:experiment}. 

Pivotal to the existence and derivation of an LR bound are the locality of Hamiltonian in space and existence of a tensor-product structure for underlying Hilbert space. Here we develop an approach in which the existence of an LR bound does not necessarily depend on a tensor-product structure for the total Hilbert space; unlike the spatial locality of the Hamiltonian (needed for the original LR bound), here we only require a representation locality. We show that if such locality condition is satisfied in a given basis of representation, the commutator of the two disjoint `local' operators in time is bounded by an LR-like bound. 

Quantum adiabatic dynamics is another context in which (relatively low) speed of a variation in the Hamiltonian of a system entails significant physical properties for the system \cite{polkovnikov:RMP,ad-review1}. This feature may conjure up an intrinsic connection between adiabaticity and LR bound. Here we provide such a bridge, and show that, in our LR-like framework, a relatively small LR speed implies adiabaticity. We also support this observation with numerical evidence.

\textit{Generalities.}---Assume a given orthonormal basis for the Hilbert space of the system. Elements (or `levels') of this basis set are ordered by assigning them consecutive labels from the set of integers. We represent each level by its label, that is, $\ket{i}$ represents level $i$ (which identifies its position in the basis labels). Next we define a \emph{block} $\mathbb{Z}$ as a subset of the basis labels, with $\abs{\mathbb{Z}}$ elements (size of the block) and $\diam(\mathbb{Z})\equiv \max_{i,i'\in \mathbb{Z}}|i-i'|$ (diameter of the block). The distance of two blocks is naturally defined as $d(\cA,\cB)\equiv\min_{i\in\cA,j\in\cB}\,|j-i|$.

\textit{An LR-like bound.}---Given a specific time-independent basis of representation $\{|i\rangle\}$, one can always rewrite the Hamiltonian in a block form as $H(t)=\sum_{\mathbb{Z}} H_{\mathbb{Z}}(t)$, where $H_{\mathbb{Z}}$ is supported on the finite block $\mathbb{Z}$ of the basis labels, and the summation is over all finite blocks. 

Suppose that $A$ and $B$ are two initially `disjoint' operators, in the sense that their associated supports $\mathbb{A}=\supp(A)$ and $\mathbb{B}=\supp(B)$ ($|\cA|$ and $|\cB|<\infty$) satisfies $\cA\cap\cB=\varnothing$ (whence also $AB=[A,B]=0$). Evolution of an operator $A$ is given in the Heisenberg picture as $A^t\equiv U^{\dag}(t,0)AU(t,0)$. The following theorem puts an upper bound on $\norm{[A^t,B]}$, where $\Vert \cdot\Vert$ is the standard operator norm.
\begin{theorem}
\label{th:LR bound}
Let the Hamiltonian $H(t)$ satisfy the following `locality' condition with respect to a given time-independent basis:
\begin{equation}
\sum_{\mathbb{Z}:~\mathbb{Z}  \cap \mathbb{P} \neq\varnothing}\abs{\mathbb{Z}} \norm{H_{\mathbb{Z}}(t)}e ^{\mu~\diam(\mathbb{Z})}\leqslant \abs{\mathbb{P}}a_\mu(t),~\forall~\mathbb{P};~|\mathbb{P}|<\infty,
\label{eq:locality-cond}
\end{equation}
where $\mu$ is a nonnegative constant, and $a_\mu(t)$ is a nonnegative integrable function of time. For any pair of disjoint operators $A$ and $B$ we shall have
\begin{equation}
\Vert[A^t,B]\Vert\leqslant 2 \min(\abs{\cA},\abs{\cB})\norm{A}\norm{B}e ^{-\mu\text{d}(\mathbb{A},\mathbb{B})}(e ^{\avg{a_\mu}_t |t|}-1),
\label{eq:LR bound}
\end{equation}
where $\cA=\supp(A)$ and $\cB=\supp(B)$ are disjoint finite blocks ($\cA \cap \cB = \varnothing$) with distance $\text{d}(\mathbb{A},\mathbb{B})$, and $\avg{a_\mu}_t= (1/t)\int_0^t a_\mu(\tau)\mathrm{d}\tau < \infty$.
\end{theorem}

We relegate the proof of the theorem to the end of the manuscript. Here, it is in order to elaborate on the locality condition (\ref{eq:locality-cond}) and some implications of the bound (\ref{eq:LR bound}). i. The locality condition (\ref{eq:locality-cond}) has been inspired by the standard (spatial) locality condition (see, e.g., Ref.~\cite{hastings:LesHouches}), where for $\forall i$ we have replaced ``site $i$" with ``level $|i\rangle$," \begin{equation}
|i\rangle\in\mathrm{basis},~~\sum_{{\mathbb{Z}}:\,{\mathbb{Z}}\ni i}|{\mathbb{Z}}|\Vert H_{{\mathbb{Z}}}(t)\Vert e^{\mu~\diam({\mathbb{Z}})}\leqslant a_{\mu}(t)<\infty.
\label{eq:hastings-local-cnd}
\end{equation}
This is equivalent to condition (\ref{eq:locality-cond}). To obtain Eq.~(\ref{eq:locality-cond}), it suffices to choose for any level $|i\rangle$, $\mathbb{P}=\{i\}$. Inversely, use the fact that $\sum_{{\mathbb{Z}}:\,{\mathbb{Z}}\cap\mathbb{P}\neq\varnothing}q_{{\mathbb{Z}}}\leqslant\sum_{i:~i\in\mathbb{P}}\,\sum_{{\mathbb{Z}}:\,{\mathbb{Z}}\ni i}q_{{\mathbb{Z}}}$, where $q_{{\mathbb{Z}}}$ is an arbitrary nonnegative quantity. Thus, a bound as $\sum_{{\mathbb{Z}}:\,{\mathbb{Z}}\ni i}q_{{\mathbb{Z}}}\leqslant a$ yields $\sum_{{\mathbb{Z}}:~{\mathbb{Z}}\cap\mathbb{P}\neq\varnothing}q_{{\mathbb{Z}}}\leqslant|\mathbb{P}| a$. 

ii. The exponential factor in Eq.~(\ref{eq:LR bound}) can be rewritten as $e^{-\mu[d(\mathbb{A},\mathbb{B})-\langle a_{\mu}\rangle_{t}|t|/\mu]}$, from whence an LR-like `speed'
\begin{equation}
V_\LR(\mu;t)\equiv \avg{a_\mu}_t /\mu,
\label{eq:V_LR-def}
\end{equation}
can be read, which captures how fast `level correlation' propagates through the dynamics. Note that $V_\LR$ is only an upper bound on the real speed, and this bound is relative to the given basis of representation one chooses. For example, if we have a time-independent Hamiltonian and choose its eigenvectors as the presentation basis, no level will propagate---whence the speed vanishes. To lower this bound on the speed, one can optimize Eq.~\eqref{eq:V_LR-def} with respect to the parameter $\mu$, which is physically related to the inverse of the `interaction' range (of the levels) in the locality condition \eqref{eq:locality-cond}. Alternatively, one can choose cleverly the ordering of the basis labels. For example, if for a pair $ij$ the value of $|H_{ij}|$ is large, we define a new set of basis labels in which these two levels are relatively closer to each other.

We remark that we could have replaced $\avg{a_\mu}_t$ with the simpler quantity $a_{\mu}^{\mathrm{m}}\equiv\sup_t a_\mu (t)$. Nevertheless, the current form of the bound has this feature that only $\avg{a_\mu}_t$ needs to exist (and be finite in the interested interval), thereby allowing for $a_{\mu}(t)$ to become large for some intermediate times---at which the Hamiltonian may even lose its locality instantaneously. 

iii.  Consider a general Hamiltonian $H=\sum_{ij}H_{ij}|i\rangle \langle j|$, written in a given orthonormal basis $\{|i\rangle\}$. This can be simply brought into a block representation by setting ${\mathbb{Z}}=\{i,j\}$; $H=\sum_{i}H_{\{i\}}+\sum_{j\neq i}H_{\{i,j\}}$ (with $\Vert H_{\{i,j\}}\Vert = |H_{ij}|$). It is straightforward to see that if $|H_{ij}|\leqslant h e^{-\mu'|i-j|}$, for some $0\leqslant h,\mu'<\infty$, in this basis this Hamiltonian is local in the sense of Eqs.~(\ref{eq:locality-cond}) or (\ref{eq:hastings-local-cnd}). Specifically, $\sum_{{\mathbb{Z}}:~{\mathbb{Z}}\cap \{i\} \neq \varnothing} |{\mathbb{Z}}|\Vert H_{{\mathbb{Z}}}\Vert e^{\mu~\diam({\mathbb{Z}})} \leqslant 4h/(1-e^{\mu-\mu'})$, which is convergent if $\mu'> \mu$. In this case, one can also find $V_{\text{LR}}=\min_{\mu}4h/[\mu(1-e^{\mu-\mu'})]$, whose minimum is attained at $\mu_{\min}=\textit{w}(e^{1+\mu'})-1$, where $\textit{w}(\textit{x})$ is the product logarithm function defined through $\textit{x}=\textit{w}(\textit{x})e^{\textit{w}(\textit{x})}$.

A Hamiltonian is called short-range local with respect to the representation basis $\{|i\rangle\}$, if for all $\mathbb{Z}$s we have $H_{\mathbb{Z};~\diam(\mathbb{Z}) > R}= 0$, for some positive $R$. In such short-range Hamiltonians, elements live around the main diagonal in the representation basis. Due to the finiteness of the set $\{\mathbb{Z}|~\mathbb{Z}\cap \mathbb{P} \neq \varnothing\}$, any short-range local Hamiltonian is also local. The theorem, however, applies to a relatively more general case than short-range local Hamiltonians. 

iv. As will be made clear later, bound (\ref{eq:LR bound}) is obtained by first calculating a bound over $\Vert A^t B\Vert$. This is in contrast to the spatial LR bound for tensor-product spaces, where the derivation differs in that one cannot simply calculate a useful bound over $\Vert A^t B\Vert$ first. Additionally, this utility in the case of our generalization allows to obtain pertinent useful bounds, e.g., for propagator of a `local" evolution. Specifically, given a fixed basis $\{|i\rangle\}$, replacing $A=\ketbra{i}{i}$ and $B=\ketbra{j}{j}$ [whence $\norm{A^t B}=\abs{\bra{j}U(t,0)\ket{i}}$] for a local Hamiltonian $H(t)$, the bound (\ref{eq:LR bound}) yields
\begin{equation}\label{eq:propagator-bound}
\abs{\bra{j}U(t,0)\ket{i}} \leqslant e ^{-\mu\,|j-i|} (e ^{\avg{a_\mu}_t |t|}-1),
\end{equation}
which implies that for a system with a local Hamiltonian, an initial state $\ket{i}$ spreads like a wave $e ^{-\mu[|j-i|-\avg{a}_t |t|/\mu]}$ in the basis space. This is an interesting general bound exhibiting how matrix elements of the exponential of a local matrix behave.

\textit{Adiabatic dynamics.}---As another important application of our LR-like bound, we shall outline an alternative approach to derivation of a form of an adiabatic condition from the LR speed. To this end, we first briefly review some relevant facts from the adiabatic theory \cite{Avron-etal:87,Ad-cond,intrinsicAQC}. Let us assume a Hamiltonian $H(t)$ whose spectrum has a band structure of $G$ (eigenprojection corresponding to, e.g., a single eigenvalue $E_G$, say, ground state) and $G^{\perp}\equiv \openone-G$ (corresponding to the rest of the spectrum), separated by a nonvanishing instantaneous gap $\Delta(t)$, such that
$H(t)=E_G(t) G(t) + G^{\perp}(t) H(t) G^{\perp}(t)$.

The dynamics of the isolated eigenprojection $G(t)$ can be described by a unitary operator $U_{\ad}(t,0;G)$ [shortly $U_{\ad}(t,0)$] with the `intertwining property' $U_{\ad}(t,0)\, G(0)\, U^{\dagger}_{\ad}(t,0) = G(t)$. One can attribute a corresponding `adiabatic Hamiltonian' $H_{\ad}(t;G)$ [shortly $H_{\ad}(t)$] to this evolution as
\begin{equation}\label{eq:H_ad}
H_{\ad}(t) = H(t) + i [\dot{G}(t),G(t)],
\end{equation}
hence, $
U_{\ad}(t,0)  =  \mathrm{Texp}[-i\int_0^t H_{\ad}(\tau)\mathrm{d}\tau]$. The adiabatic error can also be captured through \cite{Avron-etal:87,intrinsicAQC} $\delta(t) \equiv \norm{\openone - \Omega(t)}$, where $\Omega(t) \equiv U_\ad^\dagger(t,0) U(t,0)$. The evolution of $\Omega$ is in turn described by $\dot{\Omega}(t)=-i K(t)\Omega(t)$, where 
\begin{equation}
K(t)\equiv U_{\ad}^{\dagger}(t,0)[H(t)-H_{\ad}(t)]U_{\ad}(t,0).
\label{eq:kernel-omega}
\end{equation}

The adiabatic condition indicates that if $H(t)$ varies slowly enough, in the sense that
\begin{equation}
\max_{t\in[0,T]}\Vert H(t)-H_{\ad}(t)\Vert/\min_{t\in[0,T]}\Delta(t) \ll \varepsilon,
\label{eq:ad}
\end{equation}
the ground state is separated by a nonvanishing gap from the rest of the spectrum, then starting from the ground state, the final state will be $\varepsilon$-close to the ground state at time $T$ (total evolution time). In fact, since $\delta(t)\leqslant\int_0^{t}\Vert H(\tau)-H_{\ad}(\tau)\Vert\mathrm{d}\tau$, it is evident that the adiabatic condition (\ref{eq:ad}) implies a small adiabatic error. It should be remarked the traditional form of the adiabatic condition is written relatively differently as \cite{Amin:PRL}
\begin{equation}
\max_{t\in[0,T]}\Vert \dot{H}(t)\Vert/\min_{t\in[0,T]}\Delta^2(t)\ll\varepsilon,
\label{tad-cond}
\end{equation}
where dot denotes $\partial_t$. This form, however, may result in some inconsistencies \cite{Amin:PRL,Marzlin-Sanders:PRL}.

To prepare the scene to apply our theorem, we can choose the eigenbasis of the initial Hamiltonian $H(0)$ as the basis of representation, in which we label the basis levels according to their corresponding eigenvalues [e.g., $|i\rangle$ represents the eigenvector of $H(0)$ corresponding to the $i$th eigenvalue $E_i$]. This choice for level labels, however, is not guaranteed to make the LR speed as small as possible. 

Consider two operators $A$ and $B$ supported on distinct eigenspaces of $H(0)$. Specifically, $A$ (e.g., a local density matrix) is defined over the ground space, and evolves under $H(t)$ as $A^t\equiv U(t,0)AU^{\dag}(t,0)$; and $B$ (another observable or local density matrix) is defined over the excited space of $H(0)$, and evolves under $H_{\ad}(t)$ as $B_t\equiv U_{\text{ad}}(t,0)B U^{\dag}_{\text{ad}}(t,0)$. A natural object to see how adiabaticity is preserved in time is $\norm{[A^t,B_t]}=\norm{[\Omega A\Omega^{\dag},B]}$, which in fact captures how $A$ becomes mixed in the excited space. Thus, the locality condition needs to be considered for the \textit{effective} Hamiltonian $K(t)$. 

For an adiabatic evolution, an initial state in the ground space will leak only slowly into the excited space. This is, in fact, a sufficient condition for locality of $K(t)$ in the instantaneous eigenbasis of $H(t)$. But to apply our theorem, we need to have locality relative to a fixed representation basis for blocks. Interestingly, here locality of $K(t)$ relative to the eigenbasis of $H(0)$ is recovered by the  conjugation $U_{\ad}(t,0)K(t)U^{\dag}_{\ad}(t,0)$, which in turn, due to Eq.~(\ref{eq:kernel-omega}), means locality in the instantaneous eigenbasis of $H(t)$ and appropriately implies adiabaticity---for a discussion of the locality of the adiabatic Hamiltonian in the standard sense, see Ref.~\cite{osborne:PRA-ad}.  

We have $H-H_{\ad}=\sum_{{\mathbb{Z}}(t)}(H-H_{\ad})_{{\mathbb{Z}}(t)}$,
where ${\mathbb{Z}}(t)$ are the blocks in which $H(t)$ is diagonal at time $t$; thereby, from the locality condition we obtain 
\begin{multline} 
\max_{t\in[0,T]}\sum_{{\mathbb{Z}}(t):~{\mathbb{Z}}(t)\cap\mathbb{P}(t)\neq\varnothing}\norm{\bigl(H(t)-H_{\ad}(t)\bigr)_{{\mathbb{Z}}(t)}} \times \\ 
\quad e^{\mu~\diam({\mathbb{Z}}(t))} \leqslant |\mathbb{P}(t)| a^{\rm m}_\mu,~\forall \mathbb{P}(t);~|\mathbb{P}(t)|<\infty.
\label{eq:adb-loc-cnd}
\end{multline}
Since $[\dot{G},G]=G[\dot{G},G]G^{\perp} + G^{\perp}[\dot{G},G]G$, the block form of $H-H_{\ad}$ can be rewritten such that it includes only those blocks that have a nonempty intersection with the instantaneous ground space $\mathbb{G}(t)$.

Now we show that a local $K$ (in the eigenbasis of $H(0)$) and a sufficiently small, associated LR speed suffice for adiabaticity. Since the LR speed has the dimension of inverse of time [i.e., dimension of energy in the $\hbar\equiv1$ unit], it naturally should be compared with an appropriate energy scale of the system, e.g., the minimum energy gap $\Delta_{\min}\equiv \min_{t\in[0,T]}\Delta(t)$. Thus, we assume that
\begin{equation}
V_{\mathrm{LR}}^{\rm m}/\Delta_{\min}\ll \widetilde{\varepsilon},
\label{eq:ad-locality}
\end{equation}
where $V^{\mathrm{m}}_{\mathrm{LR}}\equiv a^{\rm m}_{\mu}/\mu$, and $\widetilde{\varepsilon}$ is some small number. Combining this adiabaticity condition with the locality condition (\ref{eq:adb-loc-cnd}) [with $\mathbb{P}(t)=\mathbb{G}(t)$] gives 
\begin{align}
\widetilde{\varepsilon}& \gg \frac{1}{\mu |\cG|\Delta_{\min}}\max_{t}\sum_{{\mathbb{Z}}(t)\cap\cG(t)\neq\varnothing}\norm{\bigl(H(t)-H_{\ad}(t)\bigr)_{{\mathbb{Z}}(t)}} \nonumber\\
 & \geqslant \frac{1}{\mu|\cG|\Delta_{\min}}\max_{t}\norm{ H(t)-H_{\ad}(t)}.
 \label{eq:adb-V_LR-cond}
\end{align}
Comparing this relation with Eq.~(\ref{eq:ad}) and assuming that $\varepsilon\equiv|\cG|\mu\widetilde{\varepsilon}$ to be a small number, we have obtained the very adiabatic condition. However, noting that $\Vert H(t)-H_{\ad}(t)\Vert \leqslant\Vert\dot{H}(t)\Vert/\Delta_{\min}$ \cite{intrinsicAQC}, transition from Eq.~(\ref{eq:adb-V_LR-cond}) to the traditional adiabatic condition (\ref{tad-cond}) is not necessarily rigorous, and thus must be done with appropriate mathematical care. The rederivation of the adiabatic condition, as we outlined here is, anyhow, an interesting result that bridges between the two important concepts of the LR bound and quantum adiabaticity.   

\begin{figure}[tp]
\includegraphics[width=8cm]{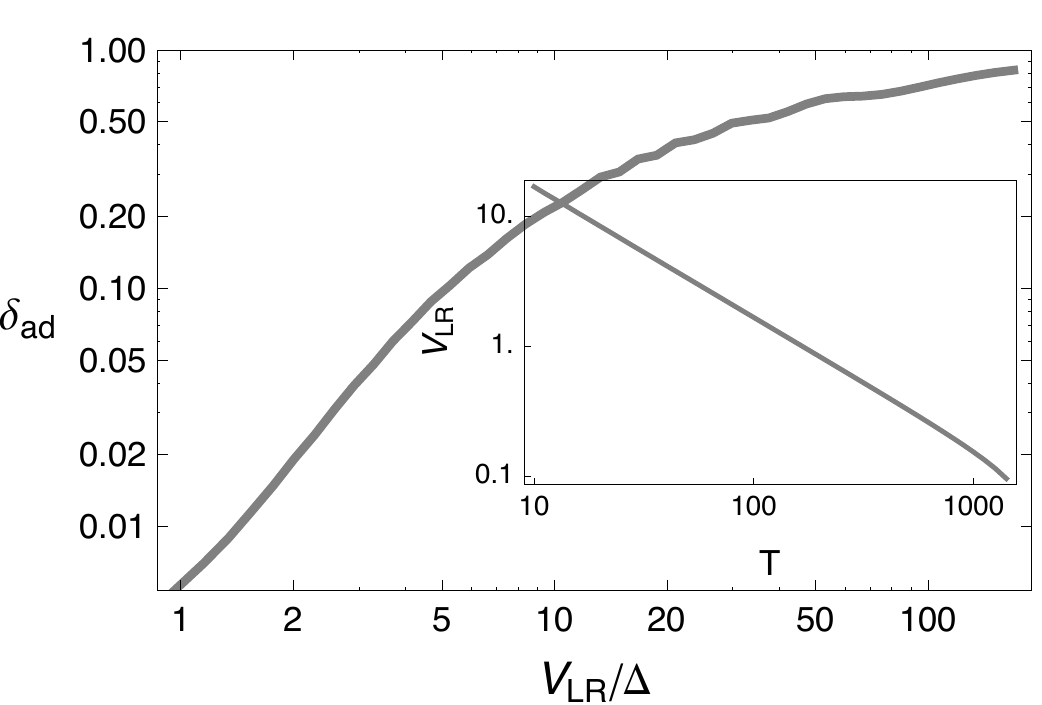}
\caption{The adiabatic error $\delta_{\ad}$ vs. the LR speed $V_{\LR}$. Inset: $V_{\LR}$ vs. the total evolution time $T$. Here, $\Delta_{\min}=0.1$ and $1.0 \leqslant\Vert H(t)\Vert \leqslant 1.8$. Both plots are in the logarithmic scale.}
\label{fig:figure}
\end{figure}

Now, we illustrate numerically the above adiabaticity and locality through a simple adiabatic process. Let us assume $H(t)=(1-t/T)H_{{\rm i}}+(t/T) H_{{\rm f}}$, where $H_{{\rm i}}=0.1 \sum_{k=0}^{10} k\ketbra{k}{k}$ and $H_{{\rm f}}=H_{\rm{i}} + (1/2)\sum_{k=0}^{9} \ketbra{k}{k+1} + \ketbra{k+1}{k}$. We note that, although this Hamiltonian is local in the initial basis, it is not local in the instantaneous basis. An LR speed can be calculated by finding the time when $\abs{\bra{E_k(t)}U(t,0)\ket{G(0)}}$ becomes larger than a specific value (in our case, $6\times 10^{-4}$). For example, by knowing this time for two different levels $\ell_{1}$ and $\ell_{2}$, one may define $V_{\LR}=(\ell_{2}-\ell_{1})/(t_{2}-t_{1})$ (of course, this calculation is meaningful when no level crossing exists in the spectrum). Figure~\ref{fig:figure} shows that the adiabatic error $\delta_{\ad}(T)\equiv1-|\langle\psi(T)|G(T)|\psi(T)\rangle|$ [where $|\psi(T)\rangle$ is the exact state of the system at time $T$] decreases when the LR speed divided by the minimum gap decreases. In addition, it exhibits that the LR speed is a decreasing function of the total time, which again can be considered as a result of adiabaticity. It is interesting that in this example, despite nonlocality of the effective Hamiltonian in the instantaneous eigenbasis of $H(t)$, still adiabaticity shows up when the LR speed is relatively small.  

textit{Proof of the theorem.}---We start from $\norm{[A^t,B]}\leqslant 2 \norm{A^t B}$, and introduce a bound for $\Vert A^t B\Vert$. The unitary invariance of the operator norm simplifies this quantity to $\norm{A^t B}=\norm{AUB}$. A Dyson expansion yields $U(t,0)=\sum_{n=0}^\infty (-i )^n \int_0^t\dd t_1 \ldots \int_0^{t_{n-1}} \dd t_n \prod\limits_{k=1}^{n} H(t_k)$. Using the block form $H(t)=\sum_{\mathbb{Z}} H_{\mathbb{Z}}(t)$, one obtains $\prod_{k=1}^n H(t_k)=\sum_{{\mathbb{Z}}_1,\ldots,{\mathbb{Z}}_n}\prod_{k=1}^n H_{{\mathbb{Z}}_k}(t_k)$. Since ${\mathbb{Z}}_i \cap {\mathbb{Z}}_j = \varnothing$ yields $H_{{\mathbb{Z}}_i} H_{{\mathbb{Z}}_j}=0$, our version of the LR bound seems simpler to prove than the original bound because in the latter (with a tensor-product structure) ${\mathbb{Z}}_i \cap {\mathbb{Z}}_j = \varnothing$ implies $[H_{{\mathbb{Z}}_i}, H_{{\mathbb{Z}}_j}] =0$, while no conclusion about $H_{{\mathbb{Z}}_i} H_{{\mathbb{Z}}_j}$ can be derived. Thus,
\begin{equation}
\prod_{k=1}^n H(t_k)=\sum_{\substack{{\mathbb{Z}}_1,\ldots,{\mathbb{Z}}_n: \\ \ch({\mathbb{Z}}_1,\ldots,{\mathbb{Z}}_n)}}
\prod_{k=1}^n H_{{\mathbb{Z}}_k}(t_k),
\label{eq:chain}
\end{equation}
where $\ch({\mathbb{Z}}_1,\ldots,{\mathbb{Z}}_n)\equiv ({\mathbb{Z}}_1\cap{\mathbb{Z}}_2\neq\varnothing) \wedge ({\mathbb{Z}}_2\cap{\mathbb{Z}}_3\neq\varnothing) \wedge \ldots \wedge ({\mathbb{Z}}_{n-1}\cap{\mathbb{Z}}_n\neq\varnothing)$ denotes a chain connecting ${\mathbb{Z}}_1$ to ${\mathbb{Z}}_n$. Now putting everything together and using the submultiplicativity of the operator norm, we obtain
\begin{multline}
\label{eq:tri-ineq}
\norm{AU(t,0)B}\leqslant\norm{A}\norm{B} \sum_{n=1}^\infty \int_0^t\dd t_1 \ldots \int_0^{t_{n-1}} \dd t_n \times\\
\bm{\sum_c}
\prod_{k=1}^n \norm{H_{{\mathbb{Z}}_k}(t_k)}, 
\end{multline}
where the summation over $n$ has begun from $n=1$ because $\cA\cap\cB=\varnothing$ implies $AB=0$, and $\bm{\sum_c}$ denotes the chain summation described in Eq.~(\ref{eq:chain}). Now we apply the locality condition (\ref{eq:locality-cond}) on each block, 
\begin{align*}
\sum_{{\mathbb{Z}}_1:~\cA\cap{\mathbb{Z}}_1\neq\varnothing}\norm{H_{{\mathbb{Z}}_1}(t_1)}e ^{\mu \, \diam({\mathbb{B}}_1)}\abs{{\mathbb{Z}}_1} &\leqslant \abs{\cA}\, a_\mu(t_1) \\
\frac{1}{\abs{{\mathbb{Z}}_1}}\sum_{{\mathbb{Z}}_2:~{\mathbb{Z}}_1\cap{\mathbb{Z}}_2\neq\varnothing}\norm{H_{{\mathbb{Z}}_2}(t_2)}e ^{\mu \,\diam({\mathbb{Z}}_2)}\abs{{\mathbb{Z}}_2} &\leqslant  a_\mu(t_2) \\
\vdots \\
\frac{1}{\abs{{\mathbb{Z}}_{n-1}}}\sum_{{\mathbb{Z}}_n:~{\mathbb{Z}}_{n-1}\cap{\mathbb{Z}}_n\neq\varnothing}\norm{H_{{\mathbb{Z}}_n}(t_n)}e ^{\mu \,\diam({\mathbb{Z}}_n)}
&\leqslant a_\mu(t_n),
\end{align*}
where in the last inequality we used the fact that $\abs{{\mathbb{B}}_n}\geqslant 1$. Combining these inequalities we get
\begin{multline}
\label{eq:combin-ineq-A}
\bm{\sum_c}
e ^{\mu\sum_{\ell} \diam({\mathbb{Z}}_{\ell})} \prod_{k=1}^n \norm{H_{{\mathbb{Z}}_k}(t_k)} 
\leqslant \abs{\cA}\prod_{k=1}^n a_\mu(t_k),
\end{multline}
in which we considered $\{{\mathbb{Z}}_n|{\mathbb{Z}}_{n-1}\cap{\mathbb{Z}}_n\neq \varnothing \wedge\,{\mathbb{Z}}_n\cap\cB\neq \varnothing\} \subseteq \{{\mathbb{Z}}_n|{\mathbb{Z}}_{n-1}\cap{\mathbb{Z}}_n\neq \varnothing\}$. If one rewrites the same series of inequalities but now begins from the $\cB$ end of the chain, the right hand side of Eq.~\eqref{eq:combin-ineq-A} becomes $ \abs{\cB}\,\prod_{k=1}^n a_\mu(t_k)$. Thus, we simply choose the lower bound by replacing $\abs{\cA}$ in the right hand side of Eq.~(\ref{eq:combin-ineq-A}) with $\min(\abs{\cA,\cB})$. 

A chain which connects $\cA$ to $\cB$ should be at least $d(\cA,\cB)$ long, i.e., $\sum_{\ell=1}^n \diam({\mathbb{B}}_{\ell})\geqslant d(\cA,\cB)$. Thus by multiplying both sides of Eq.~\eqref{eq:combin-ineq-A} by $e ^{-\mu\,d(\cA,\cB)}$, 
we obtain
\begin{multline}
\label{eq:bnd-H-prod}
\bm{\sum_c}
\prod_{k=1}^n \norm{H_{{\mathbb{B}}_k}(t_k)}
 \leqslant \min(\abs{\cA},\abs{\cB}) e ^{-\mu\,d(\cA,\cB)} \prod_{k=1}^n a_\mu(t_k). \nonumber
\end{multline}
Finally, inserting this result into Eq.~\eqref{eq:tri-ineq} and using $\int_0^t\dd t_1 \ldots \int_0^{t_{n-1}} \dd t_n\prod_{k=1}^n a_\mu(t_k) =\frac{1}{n!} {\avg{a_\mu}_t}^n t^n$ give Eq.~(\ref{eq:LR bound}). 

For the case of $t<0$, the only difference is a $(-)^n$ factor at the summation in Eq.~\eqref{eq:tri-ineq}, leading to $ |t|$ in the final formula. Clearly, the theorem also holds when $A^t\equiv U(t,0)A U^{\dag}(t,0)$, e.g., as for density matrices. In fact, because of $\Vert[UAU^\dagger,B]\Vert=\Vert[ U^\dagger B U,A]\Vert$, the bound is symmetric under $A\leftrightarrow B$. \hfill $\blacksquare$\\

\textit{Summary and outlook.}---We have developed a Lieb-Robinson-like bound---on the commutator of two observables defined on disjoint supports, one evolving dynamically, while the other one kept constant---for the case of Hamiltonians in which locality is not induced from a tensor-product structure of Hilbert space. Rather, this locality is attributed to the matrix representation in a given fixed basis, and is connected to a direct-sum structure for the Hilbert space. We have shown that this generalized locality and the ensuing bound are more conducive to some physical applications and interesting implications, e.g., on quantum propagators and adiabatic evolutions. In particular, we have demonstrated that within our framework, the adiabatic condition can be derived from an apparent locality of the matrix representation of adiabatic Hamiltonian and assuming a relatively small Lieb-Robinson speed associated to this dynamics. 

Having at hand an alternative Lieb-Robinson bound can offer a variety of relevant implications in quantum manybody systems. For example, we hope that our formulation of adiabatic condition through the Lieb-Robinson bound sheds some light on the argued role of the Anderson localization in (obstructing) adiabatic quantum computation \cite{altsh-adb-aqc}. In addition, we anticipate that the adiabatic Lieb-Robinson speed may be in an intimate correspondence with the performance of adiabatic quantum computation or algorithms; the smaller this speed is, the longer an adiabatic algorithm must take to yield an answer with some given fidelity. We hope that formalizing these expectations should be relatively natural within our framework. In a different context, investigating implications of our bound on open quantum systems with, e.g., Markovian dynamics \cite{poulin:PRL,barthel:PRL,hastings:PRL-decay}, may also offer further clues on how correlations in open systems evolve and hence affect the underlying physics. 

\textit{Acknowledgments.}---Supported by Sharif University of Technology's Office of Vice-President for Research. Helpful communications with D. A. Lidar and T. J. Osborne are acknowledged. 


\end{document}